\documentclass[pra,twocolumn,aps,superscriptaddress,showpacs,longbibliography]{revtex4-2}
\usepackage{hyperref}
\hypersetup{
	colorlinks=true,
	citecolor=blue,
	linkcolor=blue,
	urlcolor=blue}
\usepackage{graphicx}
\usepackage{amsmath}
\usepackage{physics}
\usepackage{amsfonts}
\usepackage{amssymb}
\usepackage{bm}
\usepackage{siunitx}
\usepackage{xfrac}
\usepackage[normalem]{ulem}
\usepackage{braket}
\usepackage{cases}

\usepackage{graphicx}
\usepackage{psfrag}
\usepackage{dcolumn}
\usepackage{subfigure}
\usepackage{bm}
\usepackage{float}
\usepackage{orcidlink}

\begin{document}

\title[]{Finite-temperature phase diagram and collective modes of coherently coupled Bose mixtures}

\author{Sunilkumar V\,\orcidlink{0009-0006-3839-0085}}
\email{d21082@students.iitmandi.ac.in}
\affiliation{School of Physical Sciences, Indian Institute of Technology Mandi, Mandi-175075 (H.P.), India.}

\author{Rajat\,\orcidlink{0000-0002-4882-1313}}%
 \email{rajat.19phz0009@iitrpr.ac.in}
 \affiliation{Department of Physics, Indian Institute of Technology Ropar, Rupnagar-140001, Punjab, India.}
\author{Sandeep Gautam\,\orcidlink{0000-0002-9864-6138}}%
 \email{sandeep@iitrpr.ac.in}
 \affiliation{Department of Physics, Indian Institute of Technology Ropar, Rupnagar-140001, Punjab, India.}
 \author{Arko Roy\,\orcidlink{0000-0003-4459-2880}}
 \email{arko@iitmandi.ac.in}
 \affiliation{School of Physical Sciences, Indian Institute of Technology Mandi, Mandi-175075 (H.P.), India.}
 
\begin{abstract}

We investigate the ferromagnetic–paramagnetic phase transition in coherently (Rabi) coupled Bose-Einstein condensates at zero and finite temperatures, exploring different routes to the transition by tuning the Rabi coupling or increasing the temperature at a fixed coupling.
Using the Hartree–Fock–Bogoliubov theory within the Popov approximation, we
map out the finite-temperature phase diagram of a three-dimensional homogeneous condensate. The critical line is identified from the softening of the spin gap, which at finite temperature does not strictly close but exhibits a discernible minimum. Magnetization and the spin dispersion branch reveal the progressive suppression of the ferromagnetic order with increasing temperature. In quasi-one-dimensional harmonic traps, the transition, driven by Rabi coupling, does not correspond to a sharply defined critical point. To this end, we use the softening of the spin breathing mode as an operational indicator of the transition, whose minimum shifts to lower coupling strengths with increasing temperature. Notably, the thermally driven transition causes monotonic hardening of all the spin modes. For both 
coupling and temperature-driven transition, the hybridized ``density" modes in the ferromagnetic phase acquire more density character while approaching the critical point.

\end{abstract}

\maketitle

\section{\label{intro}Introduction}

Spinor condensates provide a unique platform where superfluidity and magnetism coexist, making them ideal quantum simulators of magnetic materials~\cite{RevModPhys.85.1191, cominotti_23} and paradigmatic models of condensed matter physics~\cite{Sachdev_2011, zurek_05, farolfi_21}. 
Among the different realizations of spinor condensates, a particularly simple but rich system is the coherently (Rabi) coupled two-component Bose–Einstein condensate (BEC), realized experimentally by coupling two hyperfine states through an external radio-frequency or microwave field~\cite{Matthews_1999,farolfi_21}. In this system, the coherent coupling term mediates the interconversion between the two components, and therefore, unlike binary Bose-Bose mixtures with independently conserved populations, the particle numbers in each component are not separately conserved~\cite{stringari}. 
This qualitative difference gives rise to distinct ground-state phases and collective excitations vis-à-vis miscible and immiscible phases and their excitations in scalar binary mixtures~\cite{recati_21}.

At zero temperature, it is well established that a coherently coupled BEC with equal intra-species interactions undergoes a magnetic-like quantum phase transition from a paramagnetic to a ferromagnetic state for
strong enough inter-species interaction~\cite{PhysRevLett.105.204101, Abad2013}. Due to the lack of symmetry in the atomic interaction strengths, the system exhibits a para-to ferromagnetic transition characterized by the spontaneous $\mathbb{Z}_2$ ordering of the longitudinal spin component. By analogy with the transverse field Ising model, the transition is governed by the ratio between the Rabi coupling and the anisotropy in the intra- and inter-species interaction strengths~\cite{cominotti_23}. The excitation spectrum in the homogeneous case has been extensively studied, showing the existence of two distinct branches: a gapless density and a gapped spin branch~\cite{PhysRevA.55.2935, PhysRevA.64.013615, PhysRevA.67.023606, Abad2013}. The spin channel, in particular, exhibits rich behavior, including mode softening, hybridization, and dynamical instabilities under nonequilibrium conditions~\cite{Abad2013, PhysRevLett.113.065303,chen_19,farolfi_21,ke_25}.
The two dispersion branches of an elongated coherently coupled BEC have been experimentally measured by exciting parametric excitations, which also lead to Faraday patterns in density and longitudinal magnetization \cite{Cominotti_22}. 
Coherently coupled condensates have also been investigated in ring geometries, where the persistent currents and their decay
channels were analyzed via the Bogoliubov-de Gennes (BdG) theory~\cite{PhysRevA.93.033603}.  

Topological and non-linear excitations in coherently coupled BECs are among other research directions that have been explored. These include the rich dynamics of quantized vortices and their composite structures~\cite{PhysRevA.95.023605, tylutki_16} and the formation of vortex molecules~\cite{PhysRevLett.93.250406, eto_20,mencia_18}. In one-dimensional geometries, nonlinear wave dynamics have been probed through magnetic solitons and phase-slips~\cite{PhysRevA.95.033614,10.21468/SciPostPhys.4.3.018,farolfi_20,sanz_22,congy_16}. Related studies have explored the formation and decay of domain walls~\cite{ihara_19,gallemi_19}. More recently, connections have been made to analogue gravity~\cite{PhysRevA.96.013611,fischer_04,garay_00,
zenesini_24}. Interaction control in coherently coupled Bose–Einstein condensates has been experimentally demonstrated~\cite{sanz_22}  across a wide range of regimes, including beyond-mean-field modifications of the equation of state~\cite{Cappellaro2017,lavoine_21}, the tunability of effective three-body interactions~\cite{Bourdel_2022_tunable, Bourdel_2025}, and access to saturating-interaction nonlinearities~\cite{Bourdel_2025_saturating}.
The nonequilibrium dynamics near the ferromagnetic transition have been explored experimentally, revealing universal scaling behavior after a sudden quench across the critical point~\cite{Nicklas_2015, PhysRevA.89.033631}. The finite-temperature behavior of the equilibrium paramagnetic-ferromagnetic transition has recently been investigated using stochastic Gross–Pitaevskii simulations, revealing a linear temperature shift of the critical point and universal scaling of magnetization and fluctuations near criticality~\cite{PhysRevA.107.043301}. 

A more complex realization of coherent coupling is spin-orbit (SO) coupling, where the interplay between Rabi coupling, interactions, and momentum-dependent synthetic fields gives rise to stripe or supersolid phases~\cite{li2017stripe, *PhysRevLett.124.053605, *Chisholm_2026}. Studies of elementary excitations in SO-coupled supersolids have provided deeper insights into gapless Goldstone modes, mode hybridization, phase boundary delineation, and the emergence of roton-like minima in the dispersion~\cite{PhysRevLett.110.235302, *PhysRevLett.114.105301, *PhysRevA.95.033616, *PhysRevLett.130.156001, *rajat_25, *PhysRevA.111.023311,congy_16}. Despite this extensive body of work, the role of thermal effects on the collective modes and the interplay with harmonic confinement for Rabi-coupled BECs remains largely unexplored.
The present work aims to address this gap by providing a comprehensive study of the excitation spectrum of coherently coupled Bose gases both in the homogeneous setting and in trapped geometries, with particular emphasis on finite-temperature effects.

First, we analyze a three-dimensional homogeneous coherently coupled Bose gas at finite temperature using the Hartree–Fock–Bogoliubov (HFB) theory within the Popov approximation~\cite{PhysRevB.53.9341}. The HFB–Popov theory is a conserving and gapless finite-temperature framework, constructed to satisfy the Hugenholtz–Pines theorem associated with the broken global 
$U(1)$ symmetry, thereby guaranteeing a gapless density (phonon) Goldstone mode~\cite{PhysRev.116.489}. We map out the finite-temperature phase diagram and identify the critical line of the ferromagnetic–paramagnetic transition through the softening of the spin gap. 

Second, we extend our study to a harmonically trapped quasi-one-dimensional (quasi-1D) coherently coupled BEC, which is closer to experimental realization. We analyze how the collective excitations evolve with temperature and coupling strength, highlighting their characteristic changes across the critical point.  
To underline the effect of explicit breaking of  $\mathbb{Z}_2$ symmetry, we also examine the collective excitation of the system with unequal intraspecies interactions at zero temperature.

The paper is organised as follows: In Section~\ref{theory}, we introduce the HFB-Popov framework and derive the BdG equations to describe the collective excitation spectrum of homogeneous coherently coupled condensates. The ground-state phases at zero temperature are then analyzed within the mean-field limit, and the corresponding dispersion relations are presented in Sec.~\ref{mf0}. In Sec.~\ref{phase-diagram}, we present the finite-temperature phase diagram of coherently coupled Bose mixtures. Section~\ref{ftdr} discusses the dispersion relations at both zero and finite temperatures, and in Sec.~\ref{spingap}, we examine the temperature dependence of the spin gap and magnetization. We then turn to trapped systems in Sec.~\ref{dentrap}, where density profiles across the ferro–paramagnetic transition are analyzed using the HFB-Popov approximation. The evolution of collective modes at zero and finite temperatures is explored in Sec.~\ref{collective-modes}. Finally, the main findings of the work are summarized in Sec.~\ref{conclusions}.
\section{Homogeneous Coherently Coupled Condensates}
\subsection{HFB-Popov formalism}\label{theory}
The grand canonical Hamiltonian, in the second quantized form, describing a coherently coupled mixture of BECs under an external confinement $V({\bf r})$ is given by 
\begin{eqnarray}
    \hat{H} &=& \sum_{i=\uparrow,\downarrow} \int \,d{\rm { \bf r}} \hat\Psi^{\dag}_{i}\left[\frac{-\nabla^{2}}{2} + V({\bf r}) - \mu + \frac{g_{ii}}{2}\hat\Psi^{\dag}_{i}\hat\Psi_{i}\right]\hat\Psi_{i} \nonumber\\
    &+& g_{\uparrow \downarrow}\int \, d{\rm {\bf r}} \hat\Psi^{\dag}_ \uparrow\hat\Psi^{\dag}_ \downarrow\hat\Psi_ \uparrow\hat\Psi_ \downarrow  - \Omega\int\, d{\rm {\bf r}} (\hat\Psi^{\dag}_ \uparrow\hat\Psi_ \downarrow + \hat\Psi^{\dag}_ \downarrow\hat\Psi_ \uparrow),
    \label{hamiltonian}
\end{eqnarray}
where $\hbar = m =1$ has been considered and $\uparrow$ and $\downarrow$ denote two distinct hyperfine states. The atoms, each of mass $m$, interact via $s$-wave scattering, characterized by intra-species interaction strengths $g_{ii}$ and an inter-species interaction strength $g_{\uparrow \downarrow}$. Additionally, the two internal states are coherently coupled by an external drive that enables Rabi oscillations, i.e., a coherent transfer of atoms between the two states. Such coupling can be experimentally realized through a two-photon transition~\cite{PhysRevLett.83.3358,PhysRevLett.105.204101,farolfi_21}, described by a coupling strength $\Omega$ (assumed real and positive). 
The Bose field operators are denoted by $\hat\Psi_i$ and the chemical potential is given by $\mu$. Considering a homogeneous coherently coupled BEC with $V({\bf r})=0$, Heisenberg's equations of motion for the Bose field operators $\hat\Psi_{i}$ are 

\begin{eqnarray}
    \iota \begin{pmatrix}
  \hat{\dot\Psi}_\uparrow\\
  \hat{\dot\Psi}_\downarrow 
  \end{pmatrix}
  &=&
  \begin{pmatrix}
      \hat{h}{_\uparrow} + g_{\uparrow \uparrow}\hat\Psi^{\dag}_{\uparrow}\hat\Psi_{\uparrow} & g_{\uparrow \downarrow}\hat\Psi^{\dag}_{\downarrow}\hat\Psi_{\uparrow} \\
      g_{\uparrow \downarrow}\hat\Psi^{\dag}_{\uparrow}\hat\Psi_{\downarrow} & \hat{h}{_\downarrow} + g_{\downarrow \downarrow}\hat\Psi^{\dag}_{\downarrow}\hat\Psi_{\downarrow}
  \end{pmatrix}
  \begin{pmatrix}
  \hat\Psi_\uparrow\\
  \hat\Psi_\downarrow  
\end{pmatrix}
  \nonumber\\
  &-&
  \begin{pmatrix}
        0 & \Omega\\
        \Omega & 0
     \end{pmatrix}
     \begin{pmatrix}
  \hat\Psi_\uparrow\\
  \hat\Psi_\downarrow  
\end{pmatrix},
 \label{heisenberg}
\end{eqnarray}
where $\hat{h}{_i} = -\nabla^{2}/2 - \mu$ is the single-particle part of the Hamiltonian.
We decompose the Bose field operator as $\hat\Psi_i ({\bf r},t) = \phi_i ({\bf r}) + \delta{\hat\psi}_i({\bf r},t)$, separating the static $c-$field $\phi_i$ (condensate part) from fluctuations $\delta{\hat\psi}_i$. Substituting this into Eq.~(\ref{heisenberg}) and assuming $\langle\delta{\hat\psi}_i\rangle = \langle\delta{\hat\psi}_i^\dag\rangle = 0$, under the HFB-Popov approximation~\cite{PhysRevB.53.9341,PhysRevA.89.013617,PhysRevA.106.013304,*PhysRevA.109.033319}  we obtain the equations describing the static condensate wavefunctions $\phi_i$. These take the form of time-independent generalized Gross-Pitaevskii (GP) equations
\begin{equation} 0=
\begin{pmatrix}
 \hat{h}_{\uparrow} +L_\uparrow &  g_{\uparrow \downarrow}\langle\delta{\hat\psi}_\downarrow^\dag \delta{\hat\psi}_\uparrow\rangle -\Omega \\
 g_{\uparrow \downarrow}\langle\delta{\hat\psi}_\uparrow^\dag \delta{\hat\psi}_\downarrow\rangle -\Omega  &
 \hat{h}_{\downarrow} + L_{\downarrow}
\end{pmatrix}
\begin{pmatrix}
    \phi_\uparrow\\
    \phi_\downarrow
\end{pmatrix},
\label{order1}
\end{equation}
where $L_i =g_{i i}(n_{i} + \tilde{n}{_i})  + g_{i \bar i} n_{\bar i}$ with $\bar i = \downarrow (\uparrow)$ if $ i = \uparrow (\downarrow)$, $\tilde{n}_{i} \equiv \langle \delta\hat\psi^{\dagger}_{i}\delta\hat\psi_{i} \rangle$ denotes the non-condensate density, and $n_i = |\phi_i|^2 + \tilde{n}_i$ is the total density of the $i$-th component.
In passing, we note that the anomalous averages such as $\tilde{m}_i = \langle\delta{\hat\psi}_i \delta{\hat\psi}_i\rangle$, and $\langle \delta{\hat\psi}_i \delta{\hat\psi}_{\bar{i}} \rangle$ have been neglected throughout as discussed in Ref.~\cite{PhysRevB.53.9341}. The terms $\langle\delta{\hat\psi}_i^\dag \delta{\hat\psi}_{\bar{i}} \rangle$ represent coherence between thermal clouds and play an important role in coherently coupled BECs~\cite{PhysRevA.106.013304,PhysRevA.109.033319}. In the presence of Rabi coupling, only the total density ($\sum_{i} n_{ic}+\tilde{n}_i$) is conserved, reflecting an underlying $U(1)$ symmetry. This contrasts with the uncoupled case ($\Omega = 0$), where densities of components are individually conserved.
To derive the equation of motion for the fluctuation operator, we subtract Eq.~(\ref{order1}) from Eq.~(\ref{heisenberg}), and apply the self-consistent mean-field approximation~\cite{Proukakis_2008}
to obtain the coupled equations for the fluctuation operators. Next, we introduce the Bogoliubov ansatz through
\begin{equation}
   \delta{\hat\psi}_{i}({\bf r},t) = \sum_{j, \bf k}\left[u^{j}_{i,{\bf k}}\frac{e^{\iota {\bf k.r}}}{\sqrt{V}}\hat\alpha_{j}{e^{-\iota \omega_{j}t}} + {v^{j*}_{i,{\bf k}}\frac{e^{-\iota {\bf k.r}}}{\sqrt{V}}\hat\alpha^{\dag}_{j}e^{\iota \omega_{j}t}}\right],
   \label{bogoliubov}
\end{equation}
where $u_{i,{\bf k}}^{j}$ and $v_{i,{\bf k}}^{j}$ are the quasiparticle amplitudes corresponding to eigenenergy $\omega_j$, and $\hat{\alpha}_j$, $\hat{\alpha}_j^\dagger$ are the quasiparticle annihilation and creation operators in a box of volume $V$. Collecting terms proportional to $e^{-\iota \omega_j t}$ and $e^{\iota \omega_j t}$ yields the BdG equations given by
\begin{equation}
\omega_{j}\begin{pmatrix}
  u_{\uparrow,{\bf k}}^{j}\\ 
  v_{\uparrow,{\bf k}}^{j}\\
  u_{\downarrow,{\bf k}}^{j}\\
  v_{\downarrow,{\bf k}}^{j}
\end{pmatrix} =  \mathcal{L}\begin{pmatrix}
 u_{\uparrow,{\bf k}}^{j}\\ 
 v_{\uparrow,{\bf k}}^{j}\\
 u_{\downarrow,{\bf k}}^{j}\\
 v_{\downarrow,{\bf k}}^{j}
\end{pmatrix},
\label{matrix1}
\end{equation}
\begin{widetext}
where,
\begin{equation}
\mathcal{L} = \begin{pmatrix}
  \mathcal{L}_{\uparrow} &  g_{\uparrow \uparrow}\phi{_\uparrow}^{2} & g_{\uparrow \downarrow}(\phi_{\uparrow}\phi^{*}_{\downarrow} + \langle{\delta{\hat\psi}_{\downarrow}^{\dag}}\delta{\hat\psi_{\uparrow}}\rangle) - \Omega  & g_{\uparrow \downarrow}\phi_{\uparrow}\phi_{\downarrow}\\ 
  
   -g_{\uparrow \uparrow}\phi^{*^{2}}_{\uparrow} & \underline{\hat{\mathcal{L}}}_{\uparrow} &  -g_{\uparrow \downarrow}\phi^{*}_{\uparrow}\phi^{*}_{\downarrow}  & -g_{\uparrow \downarrow}(\phi^{*}_{\uparrow}\phi_{\downarrow} + \langle{\delta{\hat\psi}_{\downarrow}^{\dag}}\delta{\hat\psi_{\uparrow}}\rangle^{*}) + \Omega \\

   g_{\uparrow \downarrow}(\phi_{\downarrow}\phi^{*}_{\uparrow} + \langle{\delta{\hat\psi}_{\uparrow}^{\dag}}\delta{\psi_{\downarrow}}\rangle) - \Omega & g_{\uparrow \downarrow}\phi_{\downarrow}\phi_{\uparrow} & \hat{\mathcal{L}}_{\downarrow} &  g_{\downarrow \downarrow}\phi^{2}_{\downarrow}\\

   -g_{\uparrow \downarrow}\phi^{*}_{\downarrow}\phi^{*}_{\uparrow} & -g_{\uparrow \downarrow}(\phi^{*}_{\downarrow}\phi_{\uparrow} + \langle{\delta{\hat\psi}_{\uparrow}^{\dag}}\delta{\hat\psi_{\downarrow}}\rangle^{*}) + \Omega & -g_{\downarrow \downarrow}\phi^{*^{2}}_{\downarrow} & \underline{\hat{\mathcal{L}}}_{\downarrow}
   
\end{pmatrix}.
\label{matrix2}
\end{equation}
\end{widetext}
Here
$ \hat{\mathcal{L}}_{i} = \frac{1}{2}(k_{x}^2 + k_{y}^2 + k_{z}^2) -\mu + 2g_{ii}n_i + g_{\uparrow \downarrow}n_{\bar{i}}$, and 
$\underline{\hat{\mathcal{L}}}_{i}=   -\hat{\mathcal{L}}_{i}$. To obtain the eigenspectrum at $T=0$, Eq.~(\ref{matrix1}) is diagonalized to obtain $\omega_j$, $u_{i,{\bf k}}^{j}$ and $v_{i,{\bf k}}^{j}$. To account for the contribution of quantum and thermal fluctuations Eqs.~(\ref{order1}) and~(\ref{matrix1}) have to be solved self-consistently with the non-condensate density given by $ \tilde{n}_{i} = (1/V)\sum'_{j,{\bf k}}\{[|u_{i,{\bf k}}^{j}|^{2} + |v_{i,{\bf k}}^{j}|^{2}]N_{0}(\omega_j) + |v_{i,{\bf k}}^{j}|^{2}\}$ and the coherence terms are given by $\langle{\delta{\hat\psi}_{i}^{\dag}} \delta{\hat\psi_{\bar i}}\rangle  = (1/V)\sum'_{j,{\bf k}}\{[u_{i,{\bf k}}^{j*}u_{\bar{i},{\bf k}}^{j} + v_{i,{\bf k}}^{j}v_{\bar{i},{\bf k}}^{j*}]N_{0}(\omega_j) + v_{i,{\bf k}}^{j}v_{\bar{i},{\bf k}}^{j*}\}$. Here $\sum'$ denotes the summations over positive eigenvalues; $N_{0}(\omega_j) \equiv (e^{\beta \omega_{j}} -1)^{-1}$ is the Bose factor associated with the quasiparticle state with real and positive energy $\omega_j$ with $\beta$ as the inverse temperature. In the limit $T\rightarrow0$, $N_{0}(\omega_j)$ goes to zero and $\tilde{n}_{i} = (1/V)\sum'_{j,{\bf k}}|v_{i,{\bf k}}^{j}|^{2}$ accounts for the quantum depletion. 
\subsection{Ground-State Phases at $T=0$: Mean-Field Limit}
\label{mf0}
For equal intra-species interaction strengths $g_{\uparrow \uparrow} = g_{\downarrow \downarrow} =g$, the ground state corresponds to the minimum of mean-field energy density~\cite{Abad2013},
\begin{equation}
    \varepsilon = \frac{1}{4}(g+g_{\uparrow \downarrow})n^2 + \frac{1}{4}(g-g_{\uparrow \downarrow})s_z^2 - \Omega\sqrt{n^2 - s_z^2} - \mu n \, ,
    \label{energy}
\end{equation}
where \( n = n_\uparrow + n_\downarrow \) and \( s_z = n_\uparrow - n_\downarrow \) are the total and spin density, respectively. The chemical potential \( \mu \), which is the same for both components, is obtained by minimizing the energy density with respect to \( n \), which produces  
\begin{equation}
    \mu = \frac{n}{2}\left( g + g_{\uparrow \downarrow} - \frac{\Omega}{\sqrt{n_\uparrow n_\downarrow}} \right) \, .
    \label{chempot}
\end{equation}

Instead minimizing Eq.~(\ref{energy}) with respect \( s_z \) leads to the condition  
\begin{equation}
    s_z\left( g - g_{\uparrow \downarrow} + \frac{2\Omega}{\sqrt{n^2 - s_z^2}} \right) = 0 \, .
    \label{mfmag}
\end{equation}

The solutions to this equation depend on the interaction parameters. The system admits two distinct ground-state phases: a \textit{neutral paramagnetic} phase with \( s_z = 0 \), which preserves the \( U(1) \times \mathbb{Z}_2 \) symmetry, and a \textit{spin-polarized ferromagnetic} phase with  
\[
s_z = \pm n \sqrt{1 - \frac{4\Omega^2}{[(g - g_{\uparrow \downarrow})n]^2}} \, ,
\]  
which spontaneously breaks the \( \mathbb{Z}_2 \) symmetry. We emphasize that throughout this work the terms ``ferromagnetic'' and ``paramagnetic''
are defined solely in terms of the longitudinal spin component $s_z$ where the system possesses a finite condensate fraction. In particular,
the so-called paramagnetic phase corresponds to $s_z = \langle {\hat \Psi}^\dagger \hat{\sigma}_z {\hat \Psi} \rangle=0$, even though the condensed
phase possesses transverse spin polarization, $s_x = \langle {\hat \Psi}^\dagger \hat{\sigma}_x {\hat \Psi} \rangle \neq 0$  and $s_y = \langle {\hat \Psi}^\dagger \hat{\sigma}_y {\hat \Psi} \rangle$ = 0. Here ${\hat \Psi} = [{\hat \Psi_{\uparrow}}, \, {\hat\Psi_{\downarrow}}]^{T}$ with $\hat{\sigma}_{x,y,z}$ being the Pauli spin matrices. The aforementioned terminology therefore refers exclusively to the discrete
$\mathbb{Z}_2$ ordering of $s_z$, analogous to the ordering in a transverse-field Ising
model.
Introducing the critical coherent coupling parameter $\Omega_{\rm cr}(T=0)=\Omega_{\rm cr}(0)= n(g_{\uparrow \downarrow} - g)/2$, the system is paramagnetic for $\Omega > \Omega_{\rm cr}(0)$, and ferromagnetic for $\Omega < \Omega_{\rm cr}(0)$.
For a detailed discussion of coherently coupled condensate mixtures of dilute atomic gases, we refer the reader to Refs.~\cite{Abad2013,recati_21}. 
Furthermore, in the homogeneous system, the excitation spectrum is isotropic in momentum space, and the dispersion relations are~\cite{lavoine_21} 

\begin{widetext}
\begin{equation}
       \omega_{\pm} = \sqrt{C_k \pm \sqrt{C_{k}^2 - \frac{k^2}{2}\left(\frac{k^2}{2} + {\Omega}\frac{n_\uparrow + n_\downarrow}{\sqrt{n_\uparrow n_\downarrow}}\right)\left[\prod_i\left(\frac{k^2}{2} + 2gn_i + {\Omega}\sqrt{\frac{n_{\bar{i}}}{n_i}}\right) - \left(2g_{\uparrow \downarrow}\sqrt{n_\uparrow n_\downarrow} - {\Omega}\right)^2\right]}},
\label{gendspsn}
\end{equation}
\end{widetext}
with

\begin{eqnarray}
C_k &=& \frac{1}{2}\sum_{i}\left(\frac{k^2}{2} + {\Omega}\sqrt{\frac{n_{\bar{i}}}{n_i}}\right)\left(\frac{k^2}{2} + 2gn_{i} + {\Omega}\sqrt{\frac{n_{\bar{i}}}{n_{i}}}\right)\nonumber\\
&-& {\Omega}\left(2g_{\uparrow \downarrow}\sqrt{n_\uparrow n_\downarrow} - {\Omega}\right).
\end{eqnarray}
Here, $\omega_+$ and $\omega_{-}$ represent the excitation frequencies of the spin and gapless density branches, respectively. 
The spin gap, defined as $\Delta = \omega_{+}(k=0)$, is given as 
\begin{equation}
    \Delta 
           =      \sqrt { 2\Omega \sqrt{n^2 - s_{z}^2}(g - g_{\uparrow \downarrow}) + 4\Omega^2 \frac{n^2}{n^2 - s_{z}^2} }.
    \label{fsgap}
\end{equation}

 To identify whether the nature of the $j$th branch is predominantly density or spin, we compute the density modulation for each component as 
$
\delta n^j_{i,\bf k}({\bf r}) = 2\cos(\bf k \cdot \bf r)\Re\left[\phi_i \big(u_{i, \bf k}^{j*} + v_{i, \bf k}^{j}\big)\right] \quad \text{for } i = \uparrow, \downarrow,
$ and $j = \pm$
with 
 $\Re$ denoting the real part. The total density and spin modulations are $\delta n^j_{\bf k} = \delta n^j_{\uparrow,\bf k} + \delta n^j_{\downarrow, \bf k}$ and $ \delta s^j_{\bf k} = \delta n^j_{\uparrow,\bf k} - \delta n^j_{\downarrow,\bf k}$, respectively.
We introduce $
S^j_{\pm,\bf k} =  \left[(\delta n^j_{\bf k})^2 \pm (\delta s^j_{\bf k})^2\right]$ to quantify the relative contributions of density and spin fluctuations
~\cite{PhysRevA.111.033310}, and define a mode character parameter $Q^j_{\bf k} \equiv S^j_{-,\bf k}/S^j_{+,\bf k}$. 
$Q^j_{\bf k} = 1(-1)$ indicates a pure density (spin) mode. 
\begin{figure}[H]
    \centering
    \includegraphics[width=\columnwidth]{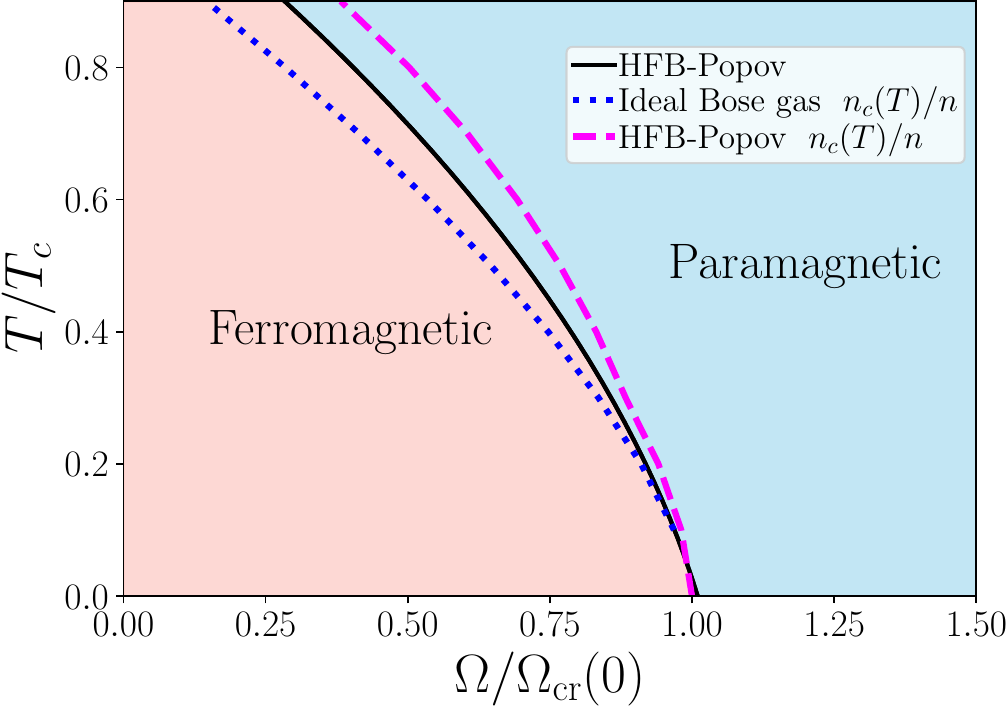}
    \caption{Finite-temperature phase diagram of a homogeneous coherently coupled BEC with $g n=1.0$ and $g_{\uparrow \downarrow}/g=1.1$. The phase boundary $\Omega_{\rm cr}(T)$ separates ferromagnetic and paramagnetic regions. HFB-Popov results (black solid line) are obtained from the ferromagnetic spin gap. Dotted lines are the ``mean-field estimates" of the phase boundary, $\Omega_{\rm cr}(T) \approx \Omega_{\rm cr}(0) n_c(T)/n$, 
    using either the condensate fraction $n_c(T)/n$ obtained from the ideal Bose gas approximation (blue dots) or that calculated from the HFB-Popov formalism (magenta dashed line).
    Figures \ref{sg}(a) and \ref{sg}(b) illustrate the typical temperature dependence of the spin gap and the magnetization $s_z/n$, respectively.}
   \label{pd}
\end{figure}
 \section{Results and Discussions}\label{results}
 \subsection{Phase Diagram}\label{phase-diagram}
  We consider a system confined to a cubic box of volume \(V= L^3 \) with periodic boundary conditions. To access its static properties at finite
temperature using the HFB–Popov formalism, we set \( L = 180 \) with 800 grid points along each $k$-space dimension and solve Eqs.~(\ref{order1}) and (\ref{matrix1}) self-consistently as described in Ref. \cite{Ritu_2025}. 
Choosing $gn$, $1/\sqrt{gn}$, and $1/gn$ as the unit of energy, length, and time, respectively, we set $gn = 1$ and $g_{\uparrow\downarrow}/g = 1.1$ for the
homogeneous system.
We consider a convergence tolerance of $10^{-3}$ for the change in thermal density during successive iterations of self-consistent calculation. 
The finite-temperature phase diagram of the system in $\Omega-T$ plane is constructed by identifying the critical line from softening of the spin gap $\Delta$ and is shown in Fig.~\ref{pd}. At $T=0$, and in the thermodynamic limit, the paramagnetic phase is characterized by a finite spin gap that closes at the ferro–paramagnetic transition and reopens upon entering the ferromagnetic phase [see Eq.~(\ref{fsgap})]. In finite-sized systems, however, the gap does not vanish strictly but reaches a discernible minimum, which we use as an operational indicator of the transition. The temperature-driven ferro-paramagnetic transition corresponds to the spontaneous
breaking of a discrete $\mathbb{Z}_2$ symmetry, for which no gapless Goldstone mode is
expected. Within the HFB--Popov approximation, the spin gap does not close at finite
temperature even in the thermodynamic limit; instead, at the vicinity of the transition, enhancement of spin susceptibility~\cite{PhysRevA.107.043301} results in the discernible minimum in the spin gap,
whose location serves as an operational indicator of the transition. Additionally, the location of the minimum of $\Delta$ was determined by performing a local quadratic fit to the data points in the vicinity of the transition. The resulting correction to the estimated critical point is smaller than the resolution set by the grid spacing in the control parameter $\Omega/\Omega_{\rm cr}(0)$. Consequently, within our numerical accuracy, the uncertainty in the phase-transition line is negligible.
With increasing temperature, the location of the critical point $\Omega_{\rm cr}(T)$
(indicated by the black solid line) shifts systematically to lower values, reflecting the gradual melting of ferromagnetic order and the corresponding amplification of the paramagnetic phase. 
At the mean-field level, the reduction of the ferromagnetic region with temperature follows from the competition between the coherent coupling $\Omega$ and the mean-field spin channel interaction term $(g_{12}-g)n_c$ being proportional to the condensate density.
In particular, 
 the softening of the spin mode yields $\Omega_{\rm cr}(T) \simeq (g_{12}-g)\,n_c(T)$, implying $\Omega_{\rm cr}(T) \approx \Omega_{\rm cr}(0)\, n_c(T)/n$.  
This can be seen explicitly from the BdG equations written in the component basis $(u_\uparrow,v_\uparrow,u_\downarrow,v_\downarrow)$ by introducing symmetric (density) and antisymmetric (spin) combinations, $u_{d/s}=(u_\uparrow \pm u_\downarrow)/\sqrt{2}$ and $v_{d/s}=(v_\uparrow \pm v_\downarrow)/\sqrt{2}$.
Under this transformation, the BdG matrix block-diagonalizes into independent density and spin sectors
in the symmetric (paramagnetic) condensate phase~\cite{Abad2013}. The spin-sector eigenvalue at $k=0$ is gapped, whose minima leads to the aforementioned criterion.
 Thermal depletion lowers $n_c(T)$, increasing the ratio $\Omega/[(g_{12}-g)\,n_c(T)]$, so that the coherent coupling increasingly dominates over the spin-channel interaction term, driving the system toward the paramagnetic phase with rising temperature.
 For a three-dimensional (3D) homogeneous, ideal two-component Bose gas, the critical temperature is given by $k_B T_c = 2\pi{[(n/2)/\zeta(3/2)]}^{2/3}$ with $n_c(T)/n = 1- (T/T_c)^{3/2}$\cite{Pethick_Smith_2008}. The corresponding variation of $\Omega_{\rm cr}(T)$ is represented by the blue dots in Fig.~\ref{pd}, which captures the low-$T/T_c$ behavior of the HFB-Popov results (black solid line) for $n_c(T)\geqslant$ 0.94. In comparison, using the condensate fraction obtained numerically from the HFB-Popov formalism in the ``mean field estimate" yields the critical line shown by the magenta dashed line in Fig.~\ref{pd}. To investigate the ferro-paramagnetic transition, we have also calculated the free energy given by\cite{hui_hu_2017},
\begin{equation}
\mathcal{F}/V = \mu \sum_{i}n_i + (E_{0} -\mu)\sum_{i}n_{ic} + \frac{k_B T}{V}\sideset{}{'}\sum_{{j,\bf{k}}}\text{ln}(1-e^{-\beta\omega_{j\bf{k}}}),
\label{free_energy}
\end{equation}
which explicitly incorporates entropy via the quasiparticle spectrum $\omega_j$. Here $E_{0}$ is the condensate energy per particle given by $E_0 = (g/2)(n_{\uparrow c}^2 + n_{\downarrow c}^2) + g_{12}n_{\uparrow c}n_{\downarrow c} - 2\Omega\sqrt{n_{\uparrow c}n_{\downarrow c}}$, and $\mu$ is the finite temperature chemical-potential obtained from the HFB-Popov calculations.  As temperature increases, the entropic contribution lowers the free energy of the system driving a transition to the thermodynamically favourable paramagnetic phase. The ferro-paramagnetic transition point coincides with the maxima of the longitudinal spin susceptibility $\chi = [\partial^2(\mathcal{F}/V)/\partial s_z^2]^{-1}$, providing an equivalent criterion for the phase boundary.

The HFB-Popov approach employed here is reliable within the condensed phase, where a finite condensate fraction is present but does not capture the physics within the normal phase. Within our numerical window, upon increasing the temperature up to $T \simeq 1.1T_c$, where the condensate fraction is reduced to $\sim 0.1$, both $s_x$ and $s_z$ decrease continuously and approach zero, while $s_y=0$ throughout. 
For $T \gg 1.1T_c$, no macroscopic condensate exists, and the system enters the normal phase that is non-magnetized with $s_z=s_x=s_y=0$. This regime is therefore qualitatively distinct from the paramagnetic condensate phase discussed in this work, which is characterized by a vanishing longitudinal magnetization $s_z=0$ but retains a finite transverse spin polarization $s_x\neq 0$
~\cite{RevModPhys.85.1191,kawaguchi_2011, kawaguchi_2012}.

\subsection{Finite temperature dispersion relations}
\label{ftdr} 
At $T=0$, our numerical results from the HFB-Popov theory are in excellent agreement with the analytical predictions of Eq.~(\ref{gendspsn}), providing an important benchmark for the validity of our computations. As shown in the top panel of
\begin{figure*}[!hbtp]
\centering
    \includegraphics[width=\textwidth]{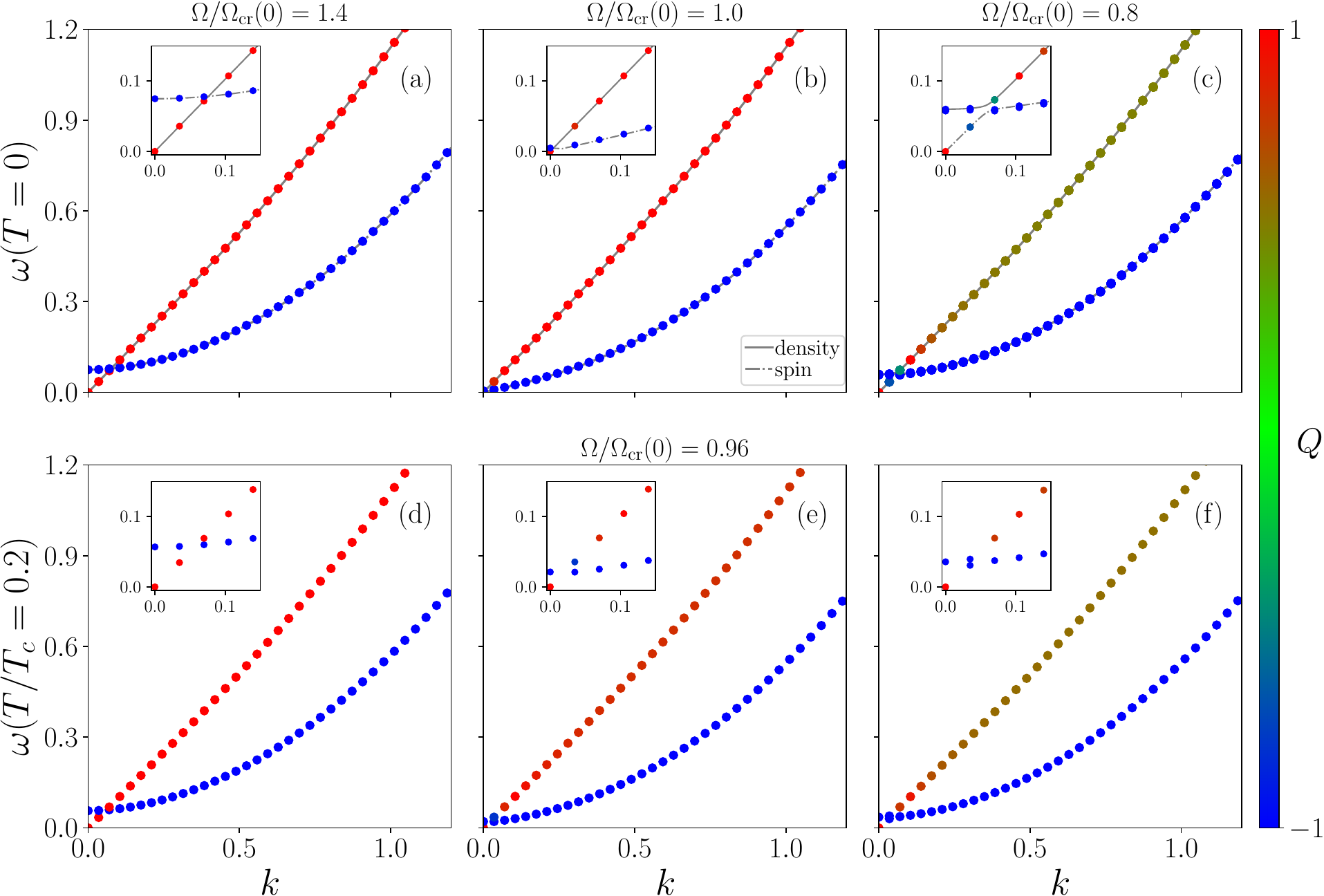}
    \caption{Dispersion relations across the ferro–paramagnetic transition. (a)–(c): Excitation spectra at $T = 0$, where solid circles correspond to the numerical solutions of Eq.~\ref{matrix1} with color denoting mode character $Q$, which has values $-1$ and $+1$ for pure spin and density mode, respectively. The dot-dashed and continuous gray lines are analytic results in
    Eq.~(\ref{gendspsn}). As the parameter \(\Omega\) decreases, the dispersion changes from (a) featuring a gapped spin branch in the paramagnetic phase, to (b) experiencing a strong suppression of the spin gap to reach its minimum at the critical point, and then (c) reopening in the ferromagnetic phase. An avoided crossing at small \(k\) indicates the presence of spin-density hybridization. (d)-(f): Excitation spectra at $T = 0.2T_c$ with similar features to those at $T = 0$ in the paramagnetic and ferromagnetic phase, but with the critical $\Omega = 0.96\Omega_{\rm cr}(0)$ accompanied by partial softening of spin gap.
    }
   \label{findisp}
\end{figure*}
Fig.~\ref{findisp}(a), for $\Omega/\Omega_{\rm cr}(0) > 1$ the system is in the paramagnetic phase. Here, the density and spin modes are clearly identified through the quantity $Q$; the density branch is gapless, while the spin branch is gapped. In addition, the two branches cross at a finite value of $k$. With decreasing $\Omega$, the spin gap gradually reduces and, at $\Omega/\Omega_{\rm cr}(0)=1$, it does not strictly close, as illustrated in Fig.~\ref{findisp}(b);  the discernible softening of the spin gap signals the onset of criticality and marks the second-order ferro-paramagnetic transition. 
  When crossing into the ferromagnetic regime for $\Omega/\Omega_{\rm cr}(0)<1$, the spin gap reopens, as seen in Fig.~\ref{findisp}(c). In this phase, avoided level crossings occur at finite $k$, accompanied by hybridization of the spin and density modes. This is evident from the values of $Q$, which no longer remain fixed at $\pm 1$ but instead interpolate between them. For instance, on the upper ($+$) branch, the modes are purely spin-like with $Q=-1$ for small $k$, but around $k\approx 0.07$ an avoided crossing occurs, and $Q$ gradually increases with increasing $k$, saturating near 0.5. Thus, the branch loses its pure spin character and represents hybrid spin--density excitations.  

The most significant change when moving from $T=0$ to $T>0$ is the shift of the critical point toward smaller values of $\Omega$. At finite temperature, the transition remains signaled by the softening of the spin gap, but the detailed structure of the modes exhibits new features. As shown in Fig.~\ref{findisp}(d), in the paramagnetic regime, the density branch remains gapless while the spin branch exhibits a reduced gap. The spin gap within the paramagnetic phase decreases with a decrease in $\Omega$ as in the case of zero temperature, but softens at the critical point. Unlike the $T=0$ case, the spectrum at the critical point, $\Omega_{\rm cr}(T) < \Omega_{\rm cr}(0)$, now exhibits avoided level crossings and hybridization of the spin and density modes [Fig.~\ref{findisp}(e)]. Entering the ferromagnetic phase, Fig.~\ref{findisp}(f), the spin gap hardens, and the hybridized nature of the excitations persists.  

 \subsection{Ferromagnetic spin gap and magnetization}
\label{spingap}
\begin{figure}[!hbtp]
\centering
    \includegraphics[width=\columnwidth]{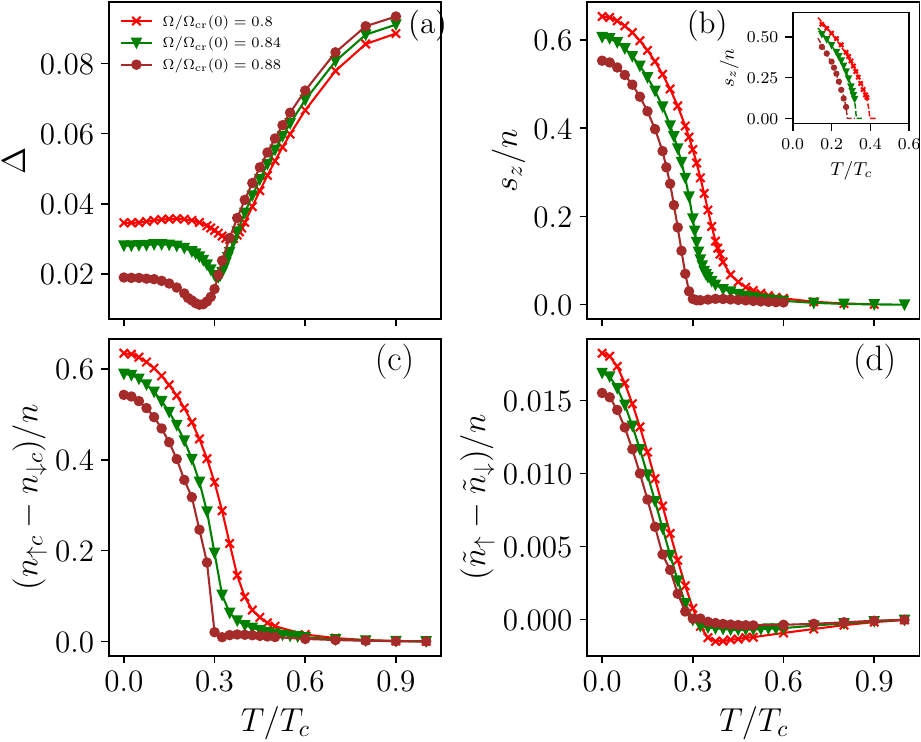}
    \caption{Temperature dependence of the ferromagnetic spin gap for different $\Omega$. The gap decreases with temperature, signaling the ferro–paramagnetic transition. Points are HFB-Popov results; lines are guides to the eye, with values extracted from Fig.~\ref{findisp}. (b) Magnetization $s_z/n$ versus temperature for the same parameters. Its rapid drop marks the transition, consistent with the spin-gap data. (Inset): $s_z/n$ fitted with $s_z \propto (1-T/T_{\rm ferro})^{\beta}$ for $T\leqslant T_{\rm ferro}$ (dashed lines), yielding $T_{\rm ferro}$ in agreement with (a) and $\beta \approx 0.5$, as expected from mean-field theory. Here $T_{\rm ferro}$ and $\beta$ are fitting parameters. 
    (c) Condensate contribution to the magnetization and (d) non-condensate contribution as functions of temperature..
    }
   \label{sg}
\end{figure}

We examine the temperature dependence of the spin gap $\Delta$ extracted from Fig.~\ref{findisp} for three representative values of $\Omega$. As shown in Fig.~\ref{sg}(a), starting with the system in the ferromagnetic phase at zero temperature, the gap decreases with temperature near the critical point, signaling the impending ferro–paramagnetic transition. Notably, the $\Delta(T)$ profiles have an asymmetry about the critical points
reflecting the distinct excitation spectra in the two phases. In the paramagnetic phase, where spontaneous magnetization is absent, the spin gap represents the energy cost of exciting spin modes and is primarily governed by the thermal energy scale. With increasing temperature, the paramagnetic state is stabilized, and the spin gap increases.

The spin gap reduction with temperature within the ferromagnetic phase reflects the loss of magnetization, as shown in Fig.~\ref{sg}(b). For each $\Omega$, the ferro–paramagnetic transition is marked by a sharp drop in $s_z/n$. Here for $\Omega$ chosen close to $\Omega_{\rm cr}(0)$, the magnetization near the critical temperature of the ferro–paramagnetic transition falls very close to zero. By contrast, for $\Omega$ values further to the left of $\Omega_{\rm cr}(0)$ display higher critical temperatures, the magnetization at the transition does not drop completely to zero but instead retains a finite value, giving the appearance of a smoother crossover~\cite{PhysRevA.107.043301}. The contrasting temperature dependence of the magnetization for different values of $\Omega$ can be traced back to the behavior of the ratio $k_B T/\Delta(T)$, which gives us a measure of the relative strength of thermal energy scale vis-\`a-vis spin gap. Close to criticality $(\Omega/\Omega_{\rm cr}(0)\rightarrow 1)$, the pronounuced softening of the spin gap leads to a rapid growth of $k_B T/\Delta(T)$, resulting in a much sharper thermal suppression of the magnetization. In contrast, deep in the ferromagnetic phase, $k_B T/\Delta(T)$ smoothly varies with $T$. Within the HFB--Popov approximation employed here, the spin gap $\Delta$ does not strictly vanish even at the phase-transition point, and finite-size effects have negligible effects on the observed trends.

Furthermore, the longitudinal magnetization can be decomposed as 
$s_z/n = (n_{\uparrow c} - n_{\downarrow c})/n + (\tilde{n}_{\uparrow} - \tilde{n}_{\downarrow})/n$, 
where the first term denotes the condensate contribution and the second term arises from 
the thermal (non-condensed) component. As shown in Fig.~\ref{sg}(c), the condensate 
contribution $(n_{\uparrow c} - n_{\downarrow c})/n$ closely follows the temperature 
dependence of the total magnetization $s_z/n$. In contrast, the thermal contribution 
$(\tilde{n}_{\uparrow} - \tilde{n}_{\downarrow})/n$ exhibits a qualitatively different 
behavior. At low $T/T_c$, when the system is fully magnetized, this quantity is positive. 
Upon increasing $T/T_c$ towards the ferro-paramagnetic transition, it decreases and 
becomes negative, with a more pronounced effect for $\Omega/\Omega_{\rm cr}(0)=0.8$ 
[Fig.~\ref{sg}(d)]. At higher temperatures, the thermal spin polarization increases again 
and ultimately vanishes, simultaneously with the condensate contribution, consistent with 
the complete loss of magnetization in the deep paramagnetic phase.

The magnetization arising from the non-condensed component can be further decomposed into 
contributions from the $j=\pm$ Bogoliubov branches,
\begin{eqnarray}
(\tilde{n}_{\uparrow}- \tilde{n}_{\downarrow}) &=& \frac{1}{V} \sideset{}{'}\sum_{j =\pm, \mathbf{k}}\Big[(|u_{\uparrow,\mathbf{k}}^{j}|^{2} - |u_{\downarrow,\mathbf{k}}^{j}|^{2})N_{0}(\omega_j) \nonumber\\
&&\qquad + (|v_{\uparrow,\mathbf{k}}^{j}|^{2} - |v_{\downarrow,\mathbf{k}}^{j}|^{2})(N_{0}(\omega_j)+1)\Big],
\label{thermal_magz}
\end{eqnarray}
where the identification of the $j=\pm$ branches with spin, spin-hybridized, and 
density modes can be inferred from $Q$ as discussed earlier. The contribution of the Bogoliubov branch to the thermal magnetization is obtained from the difference 
$|u_{\uparrow,\mathbf{k}}^{j}|^{2} - |u_{\downarrow,\mathbf{k}}^{j}|^{2}$ and the corresponding 
$v$-amplitudes, weighted by the Bose occupation $N_{0}(\omega_{j})$.
 The total 
contribution from the $j=-$ (density) branch is negative at low $T/T_c$, increases in magnitude 
with temperature, and vanishes at higher $T/T_c$. In contrast, the $j=+$ (spin-density) hybridized branch 
contributes positively at low $T/T_c$, decreases with increasing temperature, and 
also tends to zero at high $T/T_c$. When the positive $j=+$ contribution dominates over 
the negative $j=-$ contribution, the net non-condensate magnetization is positive, 
consistent with the ferromagnetic phase. Close to the ferro-paramagnetic transition, 
the $j=-$ contribution outweighs the $j=+$ branch, leading to a negative thermal 
magnetization. At higher temperatures, both branch contributions vanish, yielding 
zero non-condensate magnetization deep in the paramagnetic regime.

As shown in the inset of Fig.~\ref{sg}(b), the magnetization $s_z/n$ exhibits a thermal power-law scaling, $s_z \propto (1-T/T_{\rm ferro})^{\beta}$, with $\beta \approx 0.5$ as the mean-field critical exponent. The extracted $T_{\text{ferro}}$ agrees well with the critical temperature obtained through the minima of the spin gap from Fig.~\ref{sg}(a).

We note that experiments on coherently coupled BECs have explored both symmetric and asymmetric
interaction regimes. In particular, the experiment of Ref.~\cite{Bourdel_2022_tunable} realizes the
symmetric case $g_{\uparrow\uparrow}=g_{\downarrow\downarrow}$, for which the $\mathbb{Z}_2$ symmetry is
preserved, and a ferro-paramagnetic transition can, in principle, be defined. By contrast, in
experiments with interaction asymmetry, $g_{\uparrow\uparrow}\neq g_{\downarrow\downarrow}$
(e.g., Refs.~\cite{lavoine_21,Bourdel_2025_saturating,Leticia_2022}), the explicit breaking of the
$\mathbb{Z}_2$ symmetry removes the ferro-paramagnetic bifurcation: the ground state is always polarized,  as discussed in Sec.~\ref {collective-modes}.

\begin{figure}[!hbtp]
    \includegraphics[width=\columnwidth]{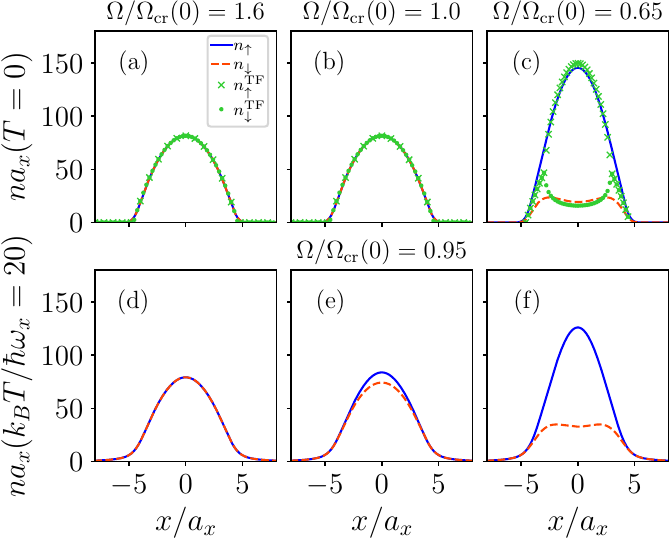}
    \caption{Total density profiles of a trapped coherently coupled condensate across the ferro-paramagnetic transition. (a)–(c) Ground-state density profiles at $T=0$, showing the transition from paramagnetic to ferromagnetic phase; GP results $(n_\uparrow,n_\downarrow)$ agree well with TF predictions $(n^{\rm TF}_\uparrow,n^{\rm TF}_\downarrow)$. (d)–(f) Finite-temperature HFB-Popov profiles, where thermal effects shift the critical point and produce elongated density tails. Dimensionless interaction parameters are $g_{\uparrow \downarrow}=1.1g$ and $g=0.06$. Here, length is expressed in units of the oscillator length $a_{x} = (\hbar/m\omega_{x})^{1/2}$.}
    \label{denprof}
\end{figure}
\section{Trapped coherently coupled Bose mixtures}
We now investigate the fate of the ferro–paramagnetic transition in a harmonically confined quasi-1D condensate. This geometry is especially relevant experimentally, as coherently coupled BECs have been typically realized in elongated traps~\cite{cominotti-23}. 
We treat the axial direction $x$ as weakly confined, with $\hbar\omega_y=\hbar\omega_z = \hbar\omega_\perp \gg kT >\hbar\omega_x$, where $\omega_{x,y,z}$ are the
trapping frequencies. In addition, to remain within the one-dimensional mean field regime, the dimensionless parameters $N a_{ii} a_{\perp}/a_x^2$ and $N a_{\uparrow\downarrow} a_{\perp}/a_x^2$ are much less than $1$, where $a_{\perp}$ and $a_x$ are the oscillator lengths along the transverse and longitudinal directions~\cite{stringari,olshanii_98,dunjko_01}. 
In this section, we use $a_x$, $\omega_x^{-1}$, $\hbar\omega_x$ as convenient units of length, time, and energy, respectively. After dimensional reduction, the effective dimensionless interaction strengths are $g=2a_{ii}(\omega_\perp/\omega_x)/a_x$ and $g_{\uparrow\downarrow}=2a_{\uparrow\downarrow}(\omega_\perp/\omega_x)/a_x$;
we consider $g_{\uparrow\downarrow} = 1.1 g$, as in the homogeneous case.
The single-particle Hamiltonian ${\hat h}_i$ in Eq. (\ref{heisenberg}) is modified to include the harmonic potential $V(x) = x^2/2$. The static condensate field is then described by the generalized GP equations (\ref{order1}), with all other symbols retaining their earlier definitions unless otherwise stated.
We use the HFB–Popov approximation to account for fluctuations and expand the fluctuation operators through the Bogoliubov transformation
\begin{equation}
    \delta{\hat\psi}_{i}(x,t) = 
    \sum_{j}\left[u_{i}^{j}(x)\hat\alpha_{j}{e^{-i\omega_{j}t}} 
    + {v^{j*}_{i}(x)\hat\alpha^{\dagger}_{j}e^{i\omega_{j}t}}\right],
    \label{bogoliubov2}
\end{equation}
where $u_i^j(x)$ and $v_i^j(x)$ denote the quasiparticle amplitudes. These functions are expanded in a linear combination of $\mathcal N$ harmonic oscillator eigenstates, and in this basis, the BdG matrix in Eq.~(\ref{matrix1}) is diagonalized to obtain the excitation spectrum. We consider $\mathcal{N}=150$, resulting in a BdG matrix of dimension $4\mathcal{N}\times 4\mathcal{N}$. The BdG matrix is generally non-Hermitian and non-symmetric, and may in principle admit complex eigenvalues. In all parameter regimes explored here, we explicitly verify that the Bogoliubov eigenfrequencies $\omega_j$ remain real throughout the self-consistent iterations, confirming the dynamical stability of the phases considered. The BdG matrix is diagonalized using the LAPACK routine \texttt{ZGEEVX}, which yields pairs of positive and negative eigenvalues. We find that the eigenfrequencies of the low-lying collective modes (in particular the spin-breathing and density-dipole modes) are converged to the order of $10^{-5}$ with respect to the basis size for the different values of Rabi coupling $\Omega$ considered in this problem at zero and finite temperatures. This demonstrates that $\mathcal{N}=150$ provides an optimal basis size for the chosen parameter set. Furthermore, thermal densities are evaluated by summing over all positive-energy Bogoliubov modes, and we verify that the Bose occupation of the highest-energy mode at the maximum temperature considered is negligible, $N_0(\omega_j)\sim 10^{-4}$. Convergence of the self-consistent scheme is assessed using the same criterion as in the homogeneous case, namely that the relative change in the local thermal densities between successive iterations falls below $10^{-3}$. We follow the numerical scheme outlined in Refs.~\cite{roy-2020, PhysRevA.89.013617} to solve Eqs.~(\ref{order1}) and ~(\ref{matrix1}) self-consistently to obtain the condensate densities $|\phi_i(x)|^2$ and the non-condensate densities $\tilde n_i(x)$.
\subsection{Ground state density profiles}\label{dentrap}
We begin by obtaining the ground-state density profiles of a quasi-\(1\)D coherently coupled Bose gas of sodium (\(^{23}\mathrm{Na}\)) atoms at zero temperature (\(T=0\)) using imaginary-time evolution implemented via the split-step Crank–Nicolson scheme. The calculations are performed for a total atom number of \(N=1000\). For sodium, the intra-species \(s\)-wave scattering lengths are taken to be equal, \(a_{\uparrow\uparrow} = a_{\downarrow\downarrow} = 54.54\,a_{B}\), where \(a_{B}\) is the Bohr radius. The atoms are confined in an anisotropic harmonic trap with transverse and axial trapping frequencies \(\omega_\perp = 2\pi \times 500~\mathrm{Hz}\) and \(\omega_x = 2\pi \times 5~\mathrm{Hz}\), respectively. These parameter choices are motivated by typical experimental conditions used in Ref.~\cite{cominotti-23}. To characterize the magnetic transition, we introduce a critical value of the Rabi coupling, defined as $\Omega_{\rm cr} (0) = (1/2)n(x=0)(g_{\uparrow \downarrow}-g)$, following Ref.~\cite{Sartori_2015} within the local density approximation. Here $n(x) = n_{\uparrow}(x) + n_{\downarrow}(x)$ is the total density, and $x=0$ corresponds to the center of the harmonic trap.
For $\Omega > \Omega_{\rm cr}(0)$, the system resides in the paramagnetic phase, whereas for $\Omega < \Omega_{\rm cr}(0)$, it favors a ferromagnetic phase. This behavior is clearly illustrated in Figs.~\ref{denprof}(a)–(c). 

In the paramagnetic regime ($\Omega > \Omega_{\rm cr}(0)$), the system remains unpolarized across the entire trap, including at the critical point in the mean-field limit. In this regime, we find very good agreement between the numerically obtained density profiles and the Thomas–Fermi (TF) analytical results of Ref.~\cite{Abad2013,Sartori_2015}, as shown in Figs.~\ref{denprof}(a) and (b). Here $n^{\rm TF}_{\uparrow,\downarrow}(x) = (({\mu^{\rm TF} + \Omega})/({g+g_{\uparrow \downarrow}}))(1- {x^2}/R^{2}_{\rm TF})$ with $\mu^{\rm TF} =[(3/8)N(g+g_{\uparrow \downarrow})]^{2/3}(1/2)^{1/3} - \Omega$ and $R^{2}_{\rm TF} = 2(\mu + \Omega)$.
On the other hand, for $\Omega < \Omega_{\rm cr}(0)$, the system may develop a polarized core while maintaining unpolarized tails~\cite{Sartori_2015}. This arises due to the inhomogeneous nature of the trapped gas: the density is highest at the trap center and decreases towards the edges. Since $\Omega_{\rm cr}$ is proportional to the local density, the condition $\Omega < \Omega_{\rm cr}(0)$ is satisfied more strongly at the trap center, where $n(x)$ is maximal. As a result, the energetically favorable solution in this regime may have a ferromagnetic (polarized) core at the center, coexisting with unpolarized wings at the edges where the density, and hence $\Omega_{\rm cr}$, is lower. Based on TF approximation, one can obtain the critical density separating the polarized core
and unpolarized tails given by $n(\pm x_c) = 2\Omega/(g_{\uparrow \downarrow}-g)$ with
$x_c =  \sqrt{2\mu + 4\Omega {g}/({g-g_{\uparrow \downarrow}})}$. For $|x|>x_c$, the density $n^{\rm TF}_{\uparrow,\downarrow} = (\mu - (1/2)x^2 + \Omega)/(g+g_{\uparrow \downarrow})$ with $\mu$ being fixed by $N$. For $|x|<x_c$, $n^{\rm TF}_{\uparrow,\downarrow} = n^{\rm TF}(1 \pm (1 - (2\Omega/(g_{\uparrow \downarrow}-g)n^{\rm TF})^2)^{1/2})$ with $n^{\rm TF} = (\mu - x^2/2)/g$ being the total density~\cite{Abad2013,Sartori_2015}. These are shown in Fig.~\ref{denprof}(c), where, for the chosen set of parameters, unlike TF estimates, numerical results do not have a well-defined unpolarized tail. 

At finite temperatures [Figs.~\ref{denprof}(d)–(f)], thermal effects elongate the density tails and shift the critical point, identified by the partial softening of the breathing mode (as will be discussed in the following subsection). A small residual magnetization persists at the transition [see Fig. \ref{denprof}(e)], and overall magnetization in the ferromagnetic phase is reduced compared to $T=0$ [cf. Figs.~\ref{denprof}(c) and (f)], demonstrating the gradual thermal destruction of ferromagnetic order.

\subsection{Collective modes}\label{collective-modes}
\subsubsection{Zero temperature}
\begin{figure}[!hbtp]
    \centering
    \includegraphics[width=\columnwidth]{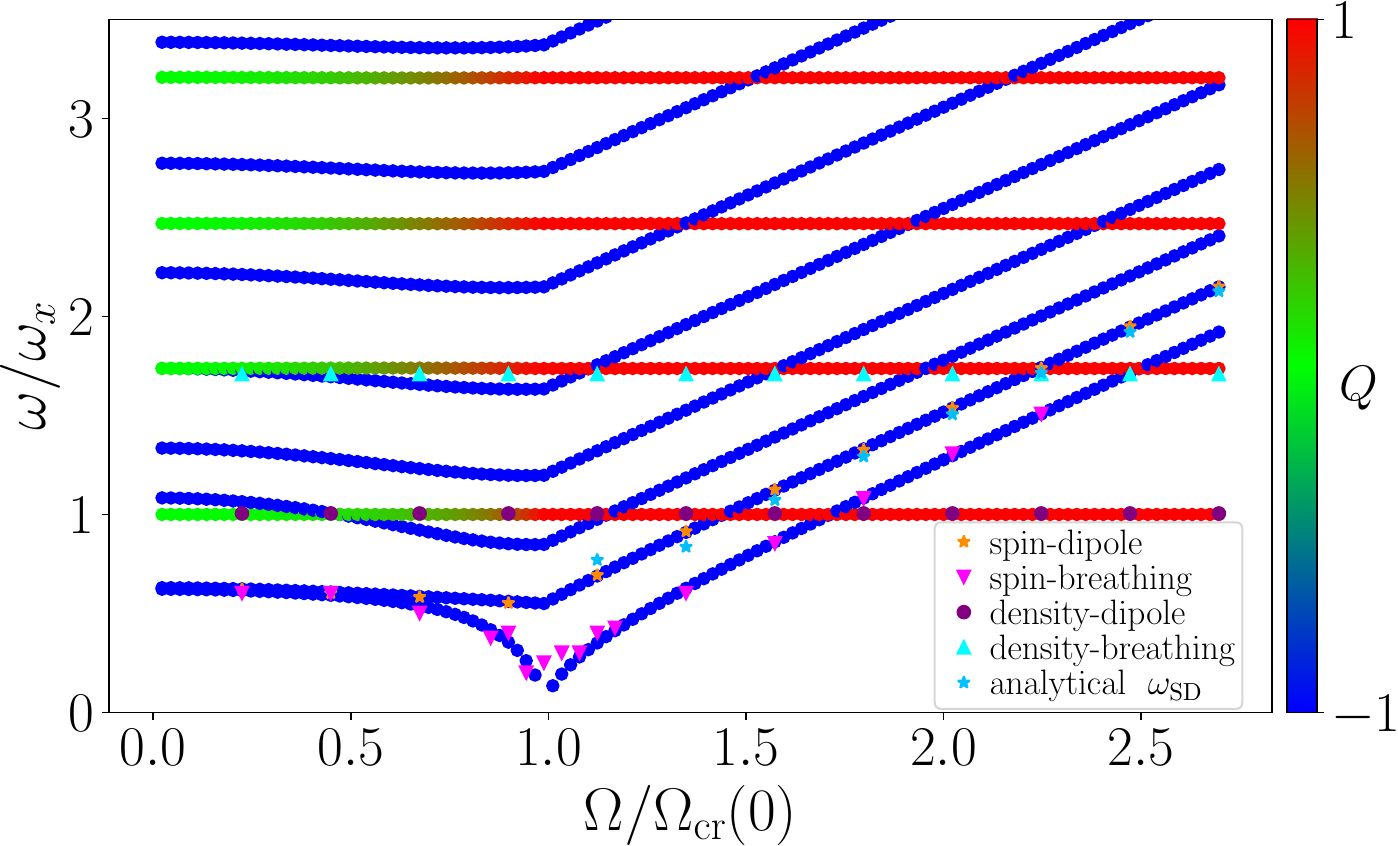}
    \caption{Excitation spectrum of a trapped coherently coupled BEC at $T=0$ from BdG calculations, compared with dynamical results (colored markers). The density and spin character of the modes is identified via the mode parameter $Q$. The softening of the spin-breathing mode marks the ferro–paramagnetic transition.}
   \label{collectivem0}
\end{figure}
Figure~\ref{collectivem0} shows the evolution of collective modes with $\Omega$ at $T=0$, providing a clear dynamical signature of the ferro–paramagnetic transition marked by the critical point $\Omega/\Omega_{\rm cr}(0)=1$. As $\Omega$ increases, the system evolves from a ferromagnetic to a paramagnetic state, with the transition indicated by the softening of the spin-breathing mode -- a signature previously demonstrated in a toroidal geometry related to the study of persistent currents~\cite{PhysRevA.93.033603}. 
In a harmonically trapped system, translational invariance is broken and the homogeneous
$k \rightarrow 0$ spin excitation is replaced by the lowest even-parity spin mode, namely the
spin-breathing (monopole) mode, which therefore constitutes the trapped analogue of the
uniform spin gap~\cite{PhysRevA.93.033603}. Although a finite trapped system does not exhibit a sharp phase
transition, the spin-breathing mode through softening encodes the finite-size remnant of the underlying
$\mathbb{Z}_2$ symmetry-breaking ferro--paramagnetic transition. 
Even at finite temperature (discussed in the next section), in the vicinity of the transition,  the spin-breathing mode remains partially softened within the HFB--Popov approximation, and develops a minimum whose location provides a practical and robust operational indicator of the transition.
Within the 
ferromagnetic phase, both the spin-breathing and spin-dipole modes decrease in energy 
and approach softening before rising again on the paramagnetic side. In contrast, the 
density-dipole and density-breathing modes remain nearly constant across both phases. In fact, the density dipole mode, 
fixed at $\omega/\omega_{x}=1$ by Kohn’s theorem~\cite{kohn_61} for all values of $\Omega$ serves as a robust benchmark for 
validating the numerical calculations. 
For small  $\Omega \lesssim 0.25$, the spin-dipole and spin-breathing modes are nearly degenerate, and the degree of degeneracy decreases with increasing $\Omega$. Furthermore, as $\Omega$ increases, the density dipole 
and breathing modes gradually recover their pure density character with $Q = S^{j}_-/S^{j}_+\rightarrow 1$ as the critical point is crossed, while the spin-dipole and spin-breathing modes retain a 
predominantly spin nature across both phases. Here $S^j_\pm = \int dx\, \left[(\delta n^j(x))^2 \pm (\delta s^j(x))^2\right] $ with the total density and spin modulations given by
$\delta n^j(x) = \delta n^j_{\uparrow}(x) + \delta n^j_{\downarrow}(x)$,  $\delta s^j(x) = \delta n^j_{\uparrow}(x) - \delta n^j_{\downarrow}(x)$, and $\delta n^j_{i}(x) = 2\,\Re\left[\phi_i(x)\,(u_{i}^{j*}(x) + v_{i}^{j}(x))\right]$~\cite{PhysRevA.108.043310}.
Furthermore on the paramagnetic side, the numerically obtained energy of the spin-dipole mode is found to be in excellent agreement with the analytical expression, $\omega_{\rm SD}^2 = \omega_{x}^2\left(\frac{g-g_{\uparrow \downarrow}}{g+g_{\uparrow \downarrow}}\right)\left[\frac{1+8\Omega n_0(g+g_{\uparrow \downarrow})/5}{1+f(\Omega/(g-g_{\uparrow \downarrow})n_0}\right]$ with $f(\alpha) = 3\alpha(1-\sqrt{1+\alpha}\, {\rm arccoth}\sqrt{(1+\alpha)})$ ~\cite{Sartori_2015}. Here $n_0$ is the total condensate density in the paramagnetic phase at $x=0$.

\begin{figure}[H]
    \centering
    \includegraphics[width=\columnwidth]{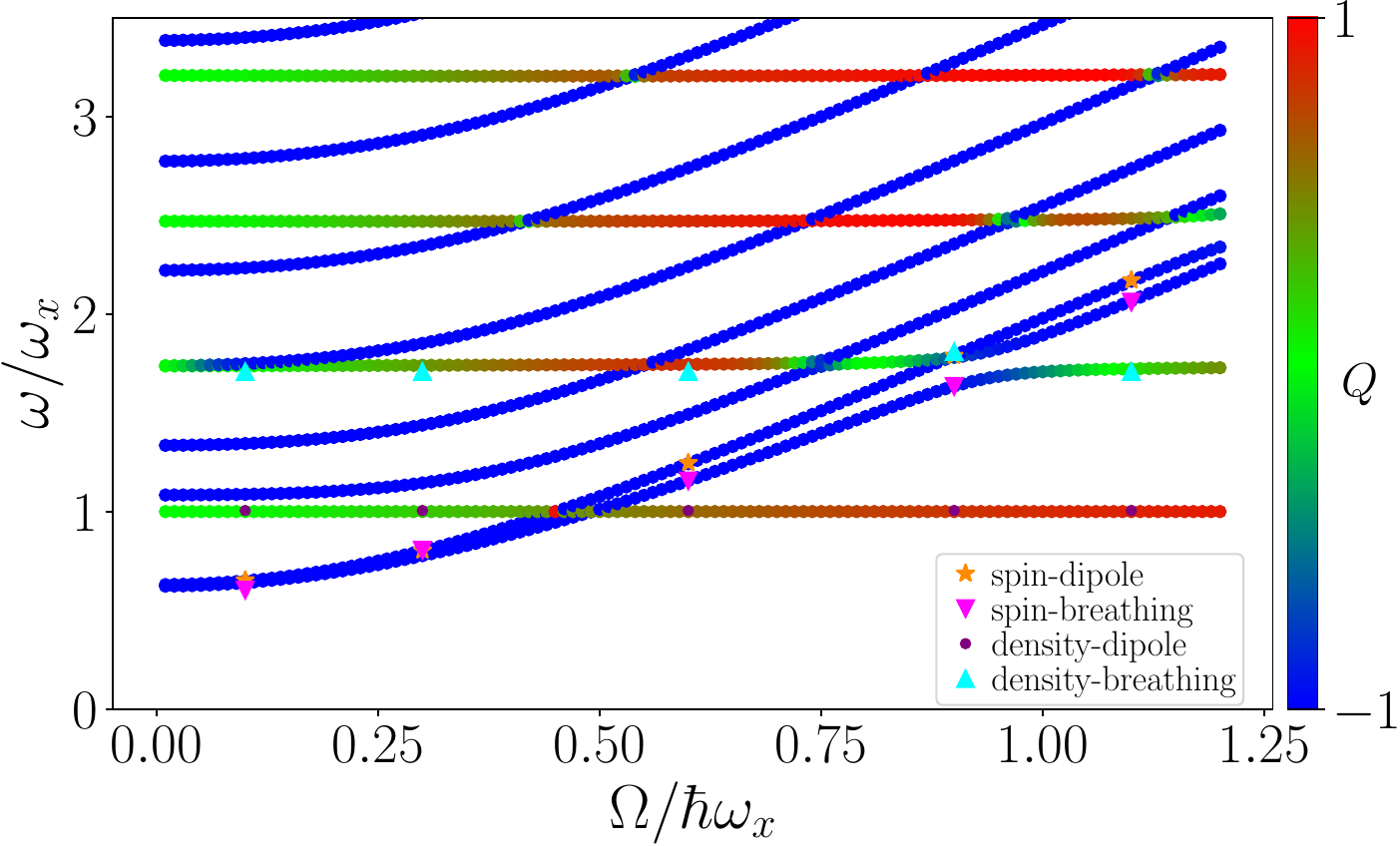}
    \caption{Excitation spectrum of a trapped coherently coupled BEC at $T=0$ with asymmetric intra-species interactions, illustrating the effect of explicit $\mathbb{Z}_2$ symmetry breaking. The parameters are $g_{\uparrow \uparrow}=0.06$, $g_{\downarrow \downarrow}=1.1g_{\uparrow \uparrow}$, and $g_{\uparrow \downarrow}=1.1g_{\uparrow \uparrow}$. Colored markers denote the frequencies extracted from real-time dynamical simulations, which are in good agreement with the BdG predictions.}
   \label{asym_exc}
\end{figure}

We also examine the case of asymmetric intra-species interactions, i.e. $g_{\uparrow \uparrow} \neq g_{\downarrow \downarrow}$, which explicitly breaks the underlying $\mathbb{Z}_2$ symmetry of the Hamiltonian. 
We choose $g_{\uparrow \uparrow}=0.06$, $g_{\downarrow \downarrow}=1.1g_{\uparrow \uparrow}$, and $g_{\uparrow \downarrow}=1.1g_{\uparrow \uparrow}$ to theoretically investigate the effects of weak interaction asymmetry,
motivated in part by experiments on SO-coupled and coherently coupled       potassium condensates where significantly larger asymmetries are explored~\cite{Chisholm_2026, sanz_22, Bourdel_2025_saturating, lavoine_21}. In this situation, the behavior of the collective modes, shown in Fig.~\ref{asym_exc}, departs from the symmetric case.
With an increase in $\Omega$, which leads to a decrease in the magnetization, density modes remain essentially constant as in the symmetric case, while the spin modes increase. The spin and density breathing modes exhibit an avoided level crossing.  
Moreover, because the interaction asymmetry enforces a small but finite magnetization even at large $\Omega$, the density modes, like the dipole mode, do not recover their pure density character as is reflected by the parameter $Q$. 

To validate the excitation spectrum, 
we numerically examine the time evolution of $\langle\hat O\rangle= \int dx \Phi^\dagger(x,t) \hat O \Phi(x,t)$ after the trapping potential is modified at $t=0$ with a 
time-independent perturbation proportional to an observable $\hat O$; here $\Phi(x,t) = [\phi_{\uparrow}(x,t), \phi_\downarrow(x,t)]^T$. 
We choose $\hat O = x, x\sigma_z, x^2$, and $ x^2\sigma_z$ to excite the dipole, spin-dipole, breathing, and spin-breathing modes, respectively.    
 In each case, the dominant oscillation frequencies of $\langle\hat O\rangle$  are compared with the BdG predictions in Figs. \ref{collectivem0} and \ref{asym_exc}. This procedure has already been implemented experimentally to probe low-lying collective excitations by modulating the trapping potential to measure density- and spin-dipole modes~\cite{PhysRevLett.77.988, bienaime_16}, and has also been examined theoretically in related studies~\cite{PhysRevA.106.013304}.
\subsubsection{Finite temperature}\label{ferro_para_zeroT}
\begin{figure}[!hbtp]
    \centering
    \includegraphics[width=\columnwidth]{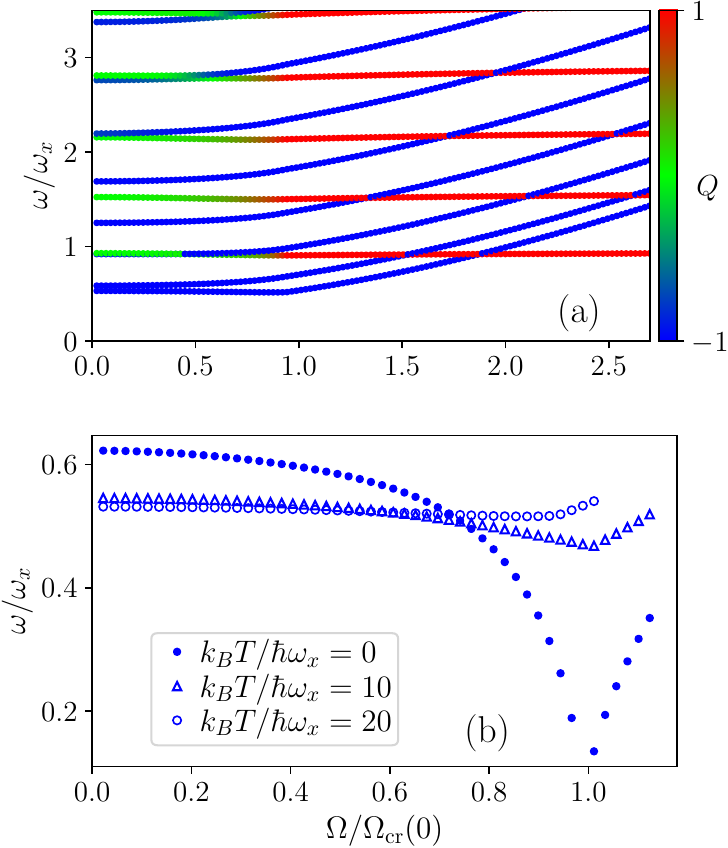}
    \caption{(a) Excitation spectrum of a trapped coherently coupled BEC at $k_BT/\hbar\omega_{x}=20$. Mode character $Q$ distinguishes spin and density branches, with the softening of the spin-breathing mode marking the ferro–paramagnetic transition [see density profiles in Fig.~\ref{denprof}(d)–(f)]. (b) 
    Spin-breathing mode from (a) with varying $\Omega$ at different temperatures, showing partial softening near the critical point for $T \neq 0$, followed by a gradual hardening in the paramagnetic phase.}
   \label{kbt60ce}
\end{figure}

In the previous section, we examined the collective excitations at zero temperature and demonstrated how the ferro–paramagnetic transition is manifested through the softening of the spin-breathing mode. We now extend this analysis to finite temperatures in order to explore how thermal and quantum fluctuations modify the collective excitation spectrum across the critical point. 
We first keep the temperature fixed, while $\Omega$ is varied as the control parameter. At finite temperature, the quantum critical region broadens~\cite{Sachdev_2011}, and as shown in Fig.~\ref{kbt60ce}(a), we illustrate partial softening of the spin-breathing mode, in contrast to the complete softening observed at $T=0$. By which we mean that the frequency of the mode at $T \neq 0$ does not reach the minimum value obtained at $T=0$. In contrast to the $T=0$ case, where the residual gap arises solely from finite-size
effects, thermal fluctuations within the HFB--Popov approximation shift the minimum of the spin-breathing mode to a higher energy,
resulting in a finite but discernible minimum that characterizes the transition. This effect is illustrated in Fig.~\ref{kbt60ce}(b), where the minimum of the spin-breathing mode progressively shifts upward in energy as the temperature increases. 

As before, the minimum of the spin-breathing mode is used to identify the critical point. The critical point has shifted to $\Omega_{\rm cr} \approx 0.95\Omega_{\rm cr}(0)$ at $k_BT/\hbar\omega_x = 20 $, where there is still non-zero magnetization as shown in Fig.~\ref{denprof}(e).  To identify the spin or density nature of the excitations, we again compute the mode character parameter $Q$.

\begin{figure}[H]
    \centering
    \includegraphics[width=\columnwidth]{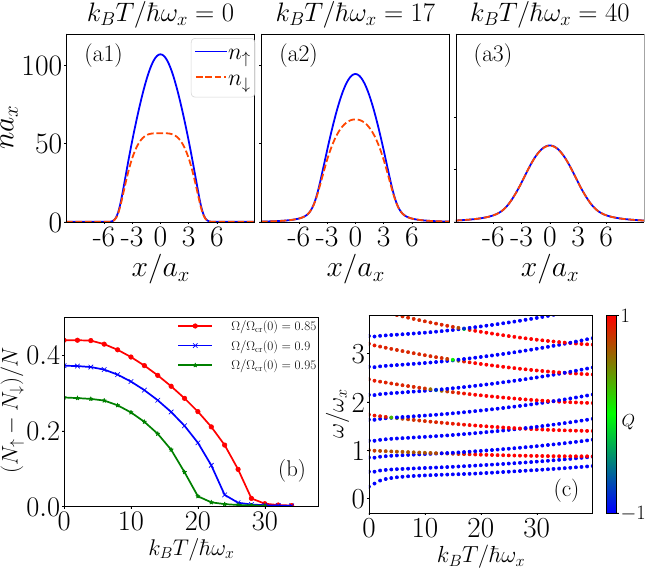}
    \caption{(a1)–(a3)Total density profiles illustrating the transformation from the ferromagnetic to the paramagnetic phase with increasing temperature for $\Omega/\Omega_{\rm cr}(0)=0.95$, (b) Temperature dependence of the global magnetization for three representative values of $\Omega/\Omega_{\rm cr}(0)$, showing its decay and eventual vanishing in the paramagnetic regime. (c) Evolution of the collective excitation spectrum for $\Omega/\Omega_{\rm cr}(0)=0.95$, highlighting the temperature-induced ferro–paramagnetic transition.}
   \label{fintempmag}
\end{figure}

 We further examine the loss of ferromagnetic order with increasing temperature at fixed $\Omega$. Starting from a ferromagnetic state at $\Omega/\Omega_{\rm cr}(0)=0.95$ and $T=0$ [Fig.~\ref{fintempmag}(a1)], the density profiles evolve with temperature [Figs.~\ref{fintempmag}(a2)-(a3)], leading to a continuous reduction of the global magnetization $(N_\uparrow - N_\downarrow)/N$. The magnetization eventually vanishes, signaling a transition to the paramagnetic phase as shown in Fig.~\ref{fintempmag}(b).
This transition is accompanied by marked changes 
in the character of collective excitation as quantified by parameter $Q$ in Fig.~\ref{fintempmag}(c). Especially, the spin–density hybridized modes gradually recover a pure density character once the critical temperature of the ferro–paramagnetic transition is crossed.
Spin modes generally increase in energy, while density modes decrease in energy with increasing temperature in both phases~\cite{PhysRevA.106.013304,*PhysRevA.109.033319}. 
 The full evolution of the excitation spectrum with temperature is summarized in Fig.~\ref{fintempmag}(c).

\section{Conclusions}\label{conclusions}

In this work, we investigated the ferromagnetic–paramagnetic phase transition in coherently coupled quantum degenerate mixtures of Bose gases by combining analytical mean-field theory with Hartree–Fock–Bogoliubov calculations within the Popov approximation. At zero temperature, we confirmed the continuous nature of the transition and benchmarked our results against mean-field predictions. Extending to finite temperature, we mapped the phase diagram of the three-dimensional homogeneous system, identifying the critical line through the softening of the ferromagnetic spin gap. 
The strong suppression of the ferromagnetic spin gap at the zero-temperature quantum critical point evolves into a partial softening at finite temperature, accompanied by a decay of magnetization from which the critical temperature can be extracted. The excitation spectra further reveal distinct density and spin excitations, as well as hybridized spin–density modes. 

For quasi-1D harmonic traps, the transition is marked by the softening of the spin-breathing mode and associated changes in the density profiles, corroborated by dynamical simulations that validate the classification of spin and density modes. At finite temperatures, 
minimum of the spin-breathing mode shifts toward lower coupling values with an increased energy. When the transition is driven by temperature, spin modes harden and density modes decrease monotonically across both the phases. The latter progressively lose their hybridized nature while approaching the critical point from within the ferromagnetic phase. Finally, introducing asymmetric intraspecies interactions causes the spin-breathing mode to harden, leading to an avoided level crossing with the density-breathing mode and a finite residual magnetization, highlighting how explicit breaking of the $\mathbb{Z}_2$ symmetry leaves a distinct imprint on the excitation spectrum.

It is worth stressing that the critical points extracted at finite temperature should be regarded as numerical outcomes rather than exact values. Our aim here is not to determine the critical line with high precision, but to provide a first quantitative characterization of the finite-temperature ferro-paramagnetic transition in a spinor superfluid, in terms of spin-gap behavior. The HFB-Popov theory, while approximate, incorporates thermal fluctuations at a reliable level and therefore offers a plausible framework for finite temperature characterization of the Rabi coupled Bose mixtures. 

Our results provide a comprehensive picture of thermal fluctuations in coherently coupled Bose mixtures, both homogeneous and trapped, and establish experimentally accessible signatures of the ferro–paramagnetic transition. Looking ahead, finite-size scaling in quasi-1D traps could clarify the universality class of the transition~\cite{sabbatini_11, williamson_16}, while quench dynamics across the critical point may reveal Kibble–Zurek scaling and defect formation~\cite{liu_25}—directions that are both theoretically appealing and experimentally accessible.

\begin{acknowledgments}
We thank Ritu, S. Patra, and Sivasankar P.M. for
several insightful discussions. A.R. acknowledges the support of the Science and Engineering Research Board (SERB),
Department of Science and Technology, Government of India, under the project
SRG/2022/000057 and IIT Mandi seed-grant funds under the project IITM/SG/AR/87.
A.R. acknowledges the National Supercomputing Mission (NSM) for providing
computing resources of PARAM Himalaya at IIT Mandi, which is implemented by
C-DAC and supported by the Ministry of Electronics and Information Technology
(MeitY) and Department of Science and Technology (DST), Government of India.
S.G. acknowledges support from the
Science and Engineering Research Board, Department of Science and Technology, Government of India, through Project
No. CRG/2021/002597. We
also thank the anonymous referees for their thorough review and valuable comments, which contributed to improving the quality of the manuscript.
\end{acknowledgments}

\section*{Data Availability}
 The data that support the findings of this study are available
 under a "Creative Commons Attribution 4.0 International License" on Zenodo under \cite{data_z}.
\bibliographystyle{apsrev4-2}
\bibliography{references}
\end{document}